\documentclass{PoS}

\usepackage{epstopdf}
\usepackage{graphicx}
\usepackage{subfigure}
\usepackage{amsmath,amssymb}
\newcommand{\nc}[1]{\newcommand{#1}}
\nc{\beqa}{\begin{eqnarray}}
\nc{\eeqa}{\end{eqnarray}}

\nc{\beq}{\begin{equation}}
\nc{\eeq}{\end{equation}}

\nc{\hmu}{\hat{\mu}}
\nc{\nn}{\nonumber}

\nc{\vecp}{\mathbf{p}}
\nc{\md}{\mathrm{d}}
\nc{\vecx}{\mathbf {x}}
\nc{\vecnull}{\mathbf {0}}

\title{Thermodynamics of heavy-light hadrons}

\ShortTitle{Thermodynamics of heavy-light hadrons}

\author{\speaker{Heng-Tong Ding} (for the BNL-Bielefeld-CCNU collaboration)\thanks{Members of the collaboration are
A. Bazavov, H.-T. Ding, P. Hegde, O. Kaczmarek, F. Karsch, E. Laermann, Y. Maezawa, Swagato Mukherjee, H. Ohno, P. Petreczky, C. Schmidt, S. Sharma, W. S\"oldner and M. Wagner.}\\
      Key Laboratory of Quark \& Lepton Physics (MOE) and Institute of
Particle Physics, \\
Central China Normal University, Wuhan 430079, China\\
        E-mail: \email{hengtong.ding@mail.ccnu.edu.cn}}

\abstract{Ratios of cumulants of conserved net charge fluctuations are sensitive to the degrees of freedom that 
are carriers of the corresponding quantum numbers in different phases of strong interaction matter. We calculate 
second and fourth order cumulants of net charm and strange fluctuations and their correlations with other conserved 
charges such as net baryon number and electric charge. Simulation are performed on $N_\tau$=6 and 8 lattices using the 
Highly Improved Staggered Quark (HISQ) action with a light to strange quark mass ratio of 1/20 and having charm 
quarks treated in the quenched approximation. Analysing appropriate ratios of these cumulants we observe that both 
open strange and charm hadrons start to get dissociated in the chiral crossover region. We provide indirect evidence for the existence of
additional, experimentally yet unobserved open charm and strange hadrons from QCD thermodynamics. This is done by comparing lattice QCD results 
to Hadron Resonance Gas (HRG) model calculations performed with a hadron spectrum as listed in the Particle Data Tables as 
well as with a spectrum predicted in the relativistic quark model and observed in lattice QCD calculations. We also discuss 
the influence of these experimentally yet unobserved states on the determination of freeze-out temperature and chemical 
potentials from heavy ion collision experiments. We found that including these additional states in the HRG model leads to a systematic 5-8 MeV decrease in
the freeze-out temperature of strange hadrons.}

\FullConference{The 32nd International Symposium on Lattice Field Theory,\\
		23-28 June, 2014\\
		Columbia University New York, NY}

\begin{document}

\section{Introduction}

Quarkonia can survive well above the transition temperature 
due to their small sizes and large binding energies as proposed by Matsui and Satz long time ago~\cite{HQ_seed} while light-quark hadrons get dissolved around transition temperature. However, the fate of heavy-light hadrons are not yet clear as e.g. strange and charm quark number susceptibilities behave quite differently from
light quark number susceptibilities. Many investigations using the Lattice QCD approach have been done on the behaviour of quarkonia and
light-quark hadrons at finite temperature~\cite{lattice}. Here we focus on the in-medium bahaviour of heavy-light quark hadrons and discuss the their manifestation
in QCD thermodynamics as well as in heavy ion collision experiments. The results presented in this proceedings have been published recently in
Ref.~\cite{MissingS, charmness} and reported in Ref.~\cite{proceedings}.

\section{Onset of the dissociation of open strange and charm hadrons}
\label{sec:deconf}

In principle all the information of hadron states are enclosed in their spectral functions.
However, it is highly non-trivial to extract hadron spectral functions from lattice QCD computations
as an analytic continuation from imaginary time to real time is needed. The spectral functions of light-quark hadrons
and heavy quarkonia have been extracted on the lattice using either the Maximum Entropy Method
and its extensions or $\chi^{2}$ fitting~\cite{lattice,Kim:2014iga}. The spectral function of heavy-light quark hadrons, however,
have not yet been investigated due to small number of data points in the temporal directions of current dynamic QCD simulations.
The properties of these heavy-light quark hadrons in medium have been studied by looking at the change of screening masses
across the transition temperature~\cite{Bazavov:2014cta}. Here we focus on the change of degree of freedom as heavy-light quarks hadrons becomes unbounded.

The properties of heavy-light quark hadrons reflect the change of the 
relevant degrees of freedom in the strong interaction medium, e.g. 
the abundance of strange hadrons is considered as one of the signals for the formation of Quark Gluon Plasma.
One obvious difference in the system when changing from hadronic phase to quark gluon plasma 
phase is that electrical charge $Q$ and baryon numbers $B$ change from integer numbers to fraction numbers.
Compared to the case in the heavy quark-antiquark system the net quantum numbers carried by the heavier quark
is not hidden in heavy-light quark hadrons. Thus the fluctuations and correlations of quantum numbers $B$ and $Q$ with strangeness
or charm allow to probe the deconfinement of carriers of strangeness and 
charm degrees of freedom, i.e. the strange and charm quarks.
These fluctuations and correlations can be calculated straightforwardly on the lattice 
defined as the derivatives of pressure with respect to 
the chemical potential of a given quark flavor. To distinguish the hadronic
and quark gluon plasma phase a featured observable needs to be constructed.

Hadron Resonance Gas (HRG) model which describes an uncorrelated hadron gas 
is a good approximation of QCD in the low temperature region.
Due to their large masses, heavy mesons 
and baryons follow Boltzmann statistics in an uncorrelated hadron gas.
 The pressure of all the strange hadrons in an uncorrelated hadron resonance gas, $P^{HRG}_S$, can be decomposed
into a mesonic, $P^{HRG}_M$, and baryonic part, $P^{HRG}_B$~\cite{strangeness}
\beq
P^{HRG}_S(\hmu_B,\hmu_S) = P^{HRG}_{|S|=1,M} \cosh(\hmu_S)  + \sum_{\ell=1}^{3}P^{HRG}_{|S|=\ell,B} \cosh(\hmu_B-\ell\hmu_S)\; .
\label{eq:P-HRG}
\eeq
Also the pressure of charmed hadron in an hadron resonance gas can be written in the same way as follows~\cite{charmness}
\beqa
P^{HRG}_C(\hmu_B,\hmu_C) &=& P^{HRG}_{|C|=1,M} \cosh(\hmu_C)  + \sum_{\ell=1}^{3}P^{HRG}_{|C|=\ell,B} \cosh(\hmu_B+\ell\hmu_C)\; \\
                                                 &=&M_C + B_{C,1} + B_{C,2} + B_{C,3}.
\eeqa
Fluctuations of conserved quantum numbers are defined as derivatives of the
pressure with respect to various chemical potentials 
$\hat{\mu}_X=\mu_X/T$, i.e.
\beq
\chi_{\rm mn}^{\rm XY} =\frac{\partial^{(m+n)} \big{(}P(\hat{\mu}_X,\hat{\mu}_Y)/T^4\big)}{\partial \hat{\mu}_X^m \partial \hat{\mu}_Y^n}\Big{|}_{\vec{\mu}=0}
\eeq
where $X,Y=B,Q,S,C$ and $\vec{\mu}=(\mu_B,\mu_Q,\mu_S,\mu_C)$ and $\chi_{0n}^{XY}\equiv\chi_n^Y$ and $\chi_{m0}^{XY}\equiv\chi_m^X$. 
It can easily observed from above relations that B-Q (S,C) correlations that differ by an even 
number of derivatives with respect to $\mu_B$ are identical.
Thus following relations for the combinations of conserved charge correlation hold~\cite{MissingS,charmness}
\beqa
\chi^{\rm BQ}_{31} / \chi^{BQ}_{11} &=& 1, \,\,\,\,\,
\chi^{\rm BS}_{31} / \chi^{BS}_{11} = 1, \,\,\,\,\,
\chi^{\rm BC}_{31} / \chi^{BC}_{11}  = 1. 
\label{eq:BQSC}
\eeqa
in an uncorrelated gas of hadrons within the classical Boltzmann approximation.
Here $\chi^{\rm BS}_{31} /\chi^{BS}_{11}$ ($\chi^{\rm BC}_{31} / \chi^{BC}_{11}$) receive contributions only from 
strange (charmed) hadrons while $\chi^{\rm BQ}_{31} / \chi^{BQ}_{11}$ 
receive contributions from all charged hadrons. Here we write the the B-C
correlations also in terms of partial pressures $M_C$ and $B_{C,i}$
\beq
\chi^{C}_m = M_C + B_{C,1} + 2^nB_{C,2}+ 3^nB_{C,3} \simeq M_C + B_{C,1},\,\,\,\,\,\,\chi_{mn}^{BC} = B_{C,1} + 2^n B_{C,2} + 3^n B_{C,3} \simeq B_{C,1}.
\eeq
The approximation in the above relations is valid due to the fact that 
baryons with charm $|C|=2,3$ have negligible contributions
to the pressure compared to those with $|C|=1$~\cite{MissingS}. 
Thus the pressure arising from charmed mesons $M_C\simeq\chi_m^C-\chi_{mn}^{BC}$. 
Together with properties show in Eq.~(\ref{eq:BQSC})
one can obtain the following relation in the hadronic phase
\beq
\chi_{nm}^{BC}\simeq\chi_{11}^{BC}\,\,~~~~ {\rm with~~n+m>0~~and~~even}.
\eeq
And the following relations
\beq
\chi_4^C = \chi_2^C,\,\,\,\,M_C = \chi_4^C - \chi_{13}^{BC} = \chi_2^C - \chi_{22}^{BC}\,\, {\rm and}\, \, \chi_{13}^{BC} = \chi_{22}^{BC}
\label{eq:BC}
\eeq
hold in the low temperature as well.

In the non-interacting case, i.e. in the high temperature limit the pressure of a charm and strange quark-antiquark gas can be written as follows in the
Boltzmann approximation
\beqa
P_{c,free}(m_c/T,\vec{\mu}/T) &=& P_{c,free}\cosh(\frac{\hmu_B}{3} + \frac{2}{3}\hmu_Q + \hmu_C), \\
P_{s,free}(m_c/T,\vec{\mu}/T) &=& P_{s,free}\cosh(\frac{\hmu_B}{3} - \frac{1}{3}\hmu_Q - \hmu_S).
\eeqa
Thus in the free case 
\beqa
\chi^{\rm BS}_{31} / \chi^{BS}_{11} =
\chi^{\rm BC}_{31} / \chi^{BC}_{11}  = 1/9,  \,\,\,\,\, \chi_{22}^{BC}/\chi_{13}^{BC}=1/3,\,\,\,\,\, \chi_4^C/\chi_2^C=\chi_{13}^{BC}/\chi_{11}^{BC}=1.
\eeqa
Thus one may expect that $\chi_4^C/\chi_2^C$ and $\chi_{13}^{BC}/\chi_{11}^{BC}$ equal to unity at all temperatures.

Knowing behaviour of these observables mentioned above in the low temperature and high temperature region, we performed
lattice QCD simulations using Highly Improved Staggered fermions on $32^3\times8$ and $24^3\times6$ lattices in the 
vicinity of chiral crossover temperature region and above.
The strange quark mass $m_s$ is fixed to its physical value and the masses of degenerate up and down quarks have
been fixed to $m_l=m_s/20$ to have a pion mass about 160 MeV in the continuum limit. The charm quark is treated in the quenched
approximation and its mass is determined
at zero temperature by calculating the spin-averaged charmonium mass $\frac{1}{4}(m_{\eta_c}+3m_{J/\psi})$. The details 
of simulations can be found in Refs.~\cite{MissingS,charmness}.

\begin{figure}[!th]
\begin{center} 
\includegraphics[width=0.35\textwidth]{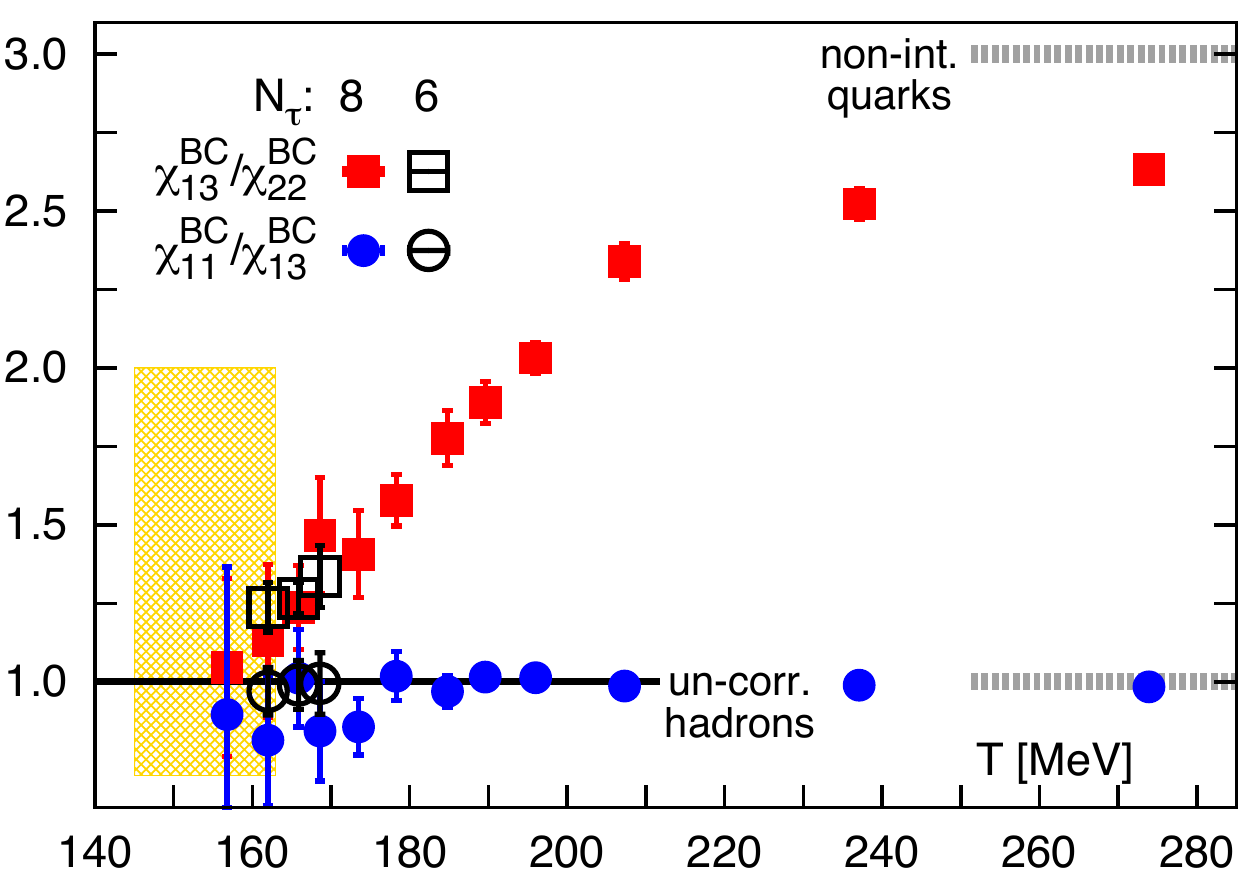}\includegraphics[width=0.35\textwidth]{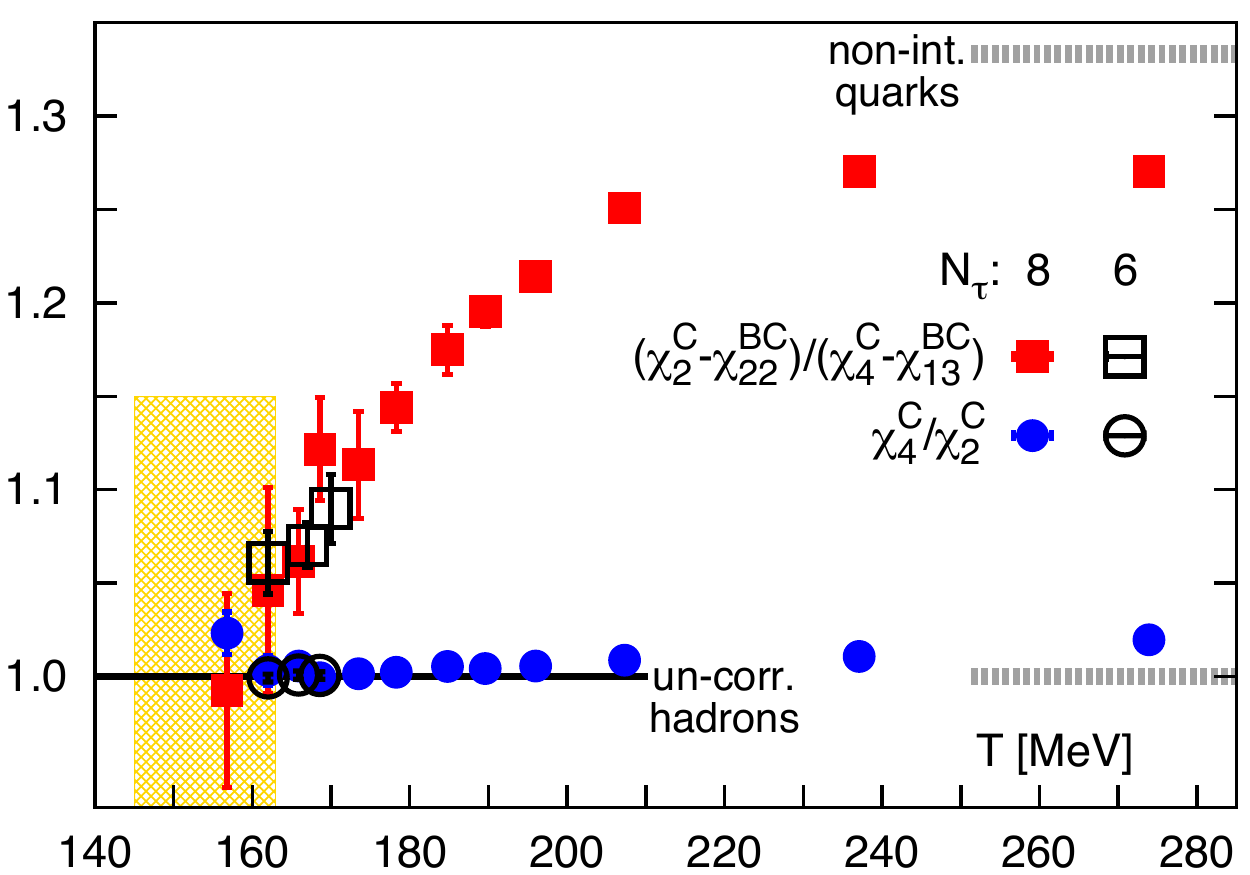}
\caption{Temperature dependences of variables discussed in relations (2.5) and (2.8).
Yellow bands shown in these two plots represent the temperature
window of the chiral crossover $T_c=154\pm 9$ MeV~\cite{Tc}.}
\label{fig:BC}
\end{center}
\end{figure}

In Fig.~\ref{fig:BC} we show the temperature dependences
of observables discussed in relations~(\ref{eq:BQSC}) and~(\ref{eq:BC}) , i.e. $\chi_{13}^{BC}/\chi_{22}^{BC}$, $\chi_{11}^{BC}/\chi_{13}^{BC}$,
$(\chi_4^C - \chi_{13}^{BC})/(\chi_2^C - \chi_{22}^{BC})$ and $\chi_4^C/\chi_2^C$.
We found that $\chi_{11}^{BC}/\chi_{13}^{BC}$ and $\chi_4^C/\chi_2^C$ are very close to unity 
near and above the chiral crossover temperature region as expected. 
The proximity of 
$\chi_{13}^{BC}/\chi_{22}^{BC}$ and $(\chi_4^C - \chi_{13}^{BC})/(\chi_2^C - \chi_{22}^{BC})$ 
to unity in the vicinity of the chiral crossover temperature region also testify the validity of the relations~(\ref{eq:BQSC}) and~(\ref{eq:BC}).
\begin{figure}[!th]
\begin{center} 
\includegraphics[width=0.36\textwidth]{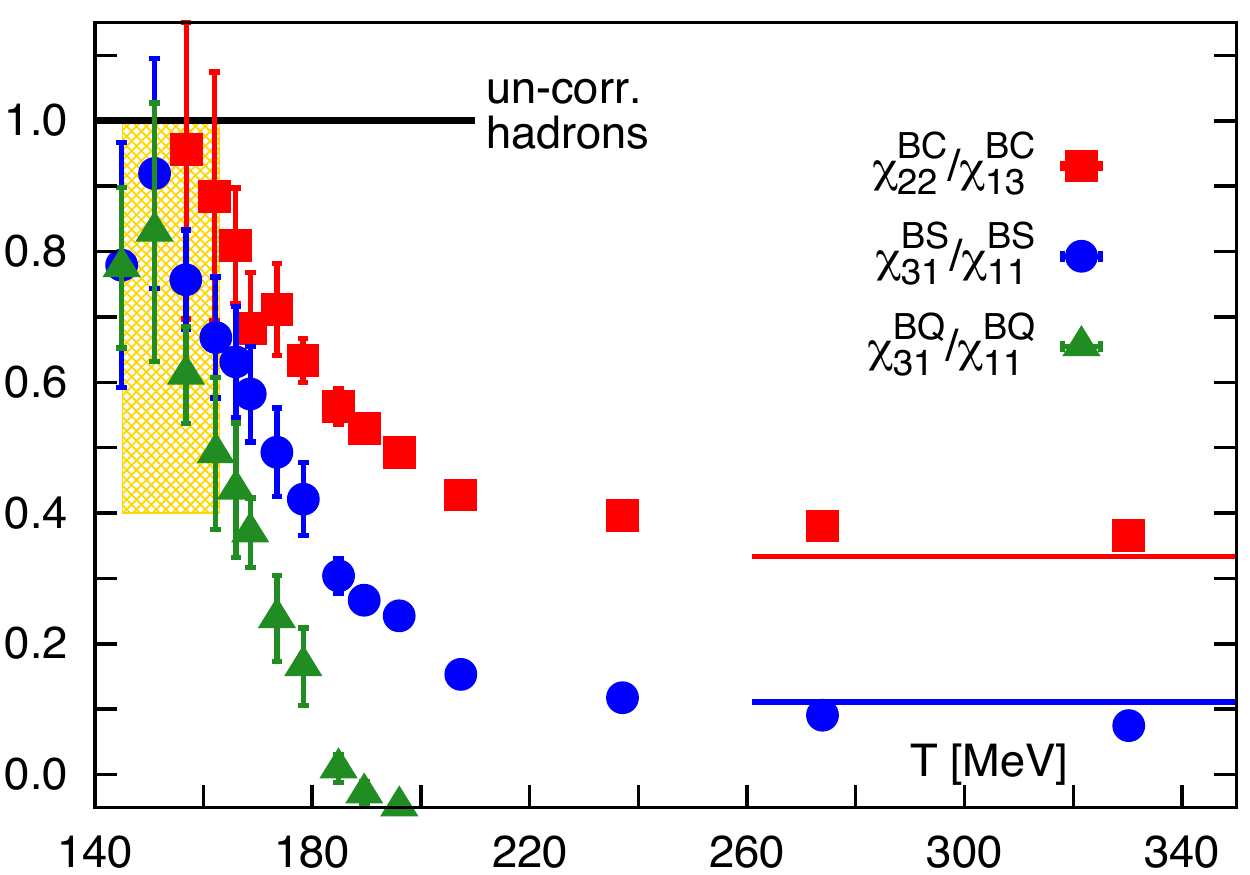}
\caption{Onset of the dissociation of open charm and open strange hadrons 
happen in the chiral crossover temperature region~\cite{charmness}. Yellow band denotes the chiral crossover temperature region.}
\label{fig:deconf}
\end{center}
\end{figure}

We show the ratios of baryon-charm ($\chi_{22}^{BC}/\chi_{13}^{BC}$),
baryon-strangeness ($\chi_{31}^{BS}/\chi_{11}^{BS}$) and baryon-electric charge ($\chi_{31}^{BQ}/\chi_{11}^{BQ}$) correlations
in Fig.~\ref{fig:deconf}. These correlations should be equal to unity as long as 
an uncorrelated hadron resonance gas model provides an appropriate description
of the thermodynamics of the medium~\footnote{In principle the correlation $\chi_{31}^{BC}/\chi_{11}^{BC}$ shows the similar behaviour as
$\chi_{22}^{BC}/\chi_{13}^{BC}$, however, the noise to signal ratio of this quantity is too high to extract any useful information.}. 
It is obvious from Fig.~\ref{fig:deconf} that
all three quantities start to deviate from unity
in the chiral crossover region.
This indicates that a description in terms of a HRG model breaks down for 
baryonic correlations involving light, strange and charm quarks, i.e. open charm/strange hadrons start to 
dissociate in or just above the chiral crossover region.

\section{Thermodynamic contributions from unobserved open strange and charm hadrons near QCD transition}

An Hadron Resonance Gas model that approximates QCD in principle should include all hadron states 
that are predicted by Quantum Chromodynamics. 
However, not all the hadron states that are predicted in the relativistic Quark 
Model (QM) and lattice QCD calculations \cite{Padmanath:2013bla} have been observed
in the experiments~\cite{QM}  and thus are listed in the particle data tables.
It is then interesting to see whether 
these additional states not listed in the particle data table have any significant imprints in QCD 
thermodynamics~\cite{MissingS,charmness}.

\begin{figure}[!th]
\begin{center} 
\includegraphics[width=0.3\textwidth,height=4.2cm]{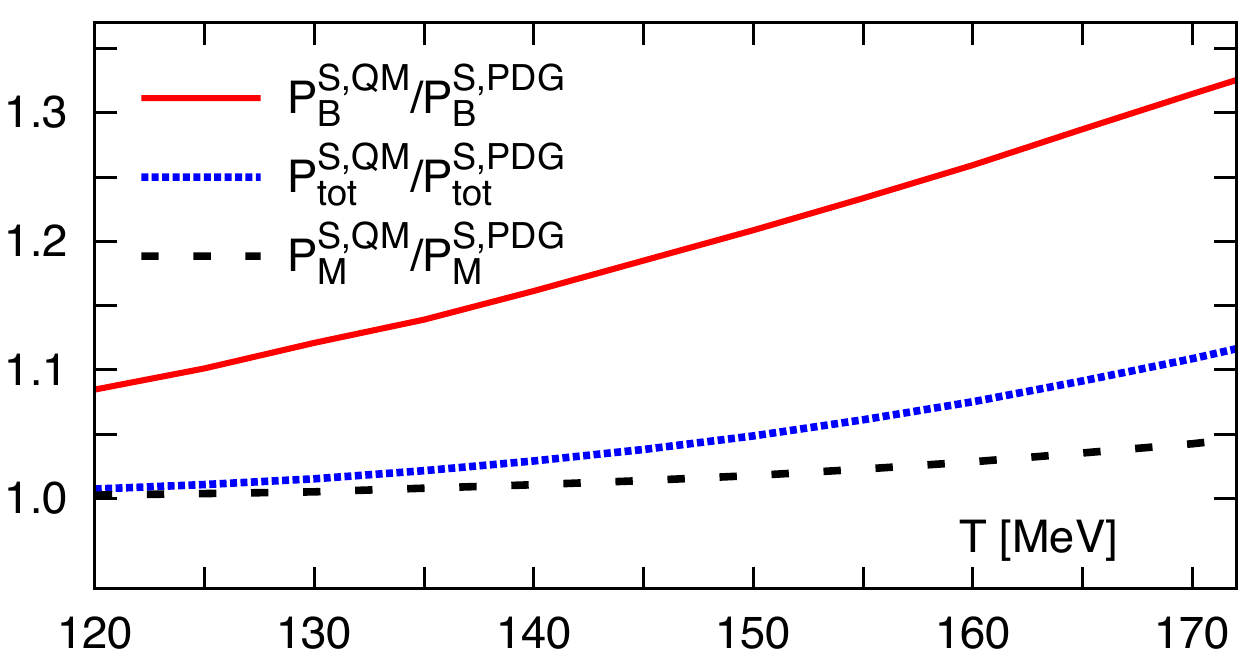}\includegraphics[width=0.3\textwidth]{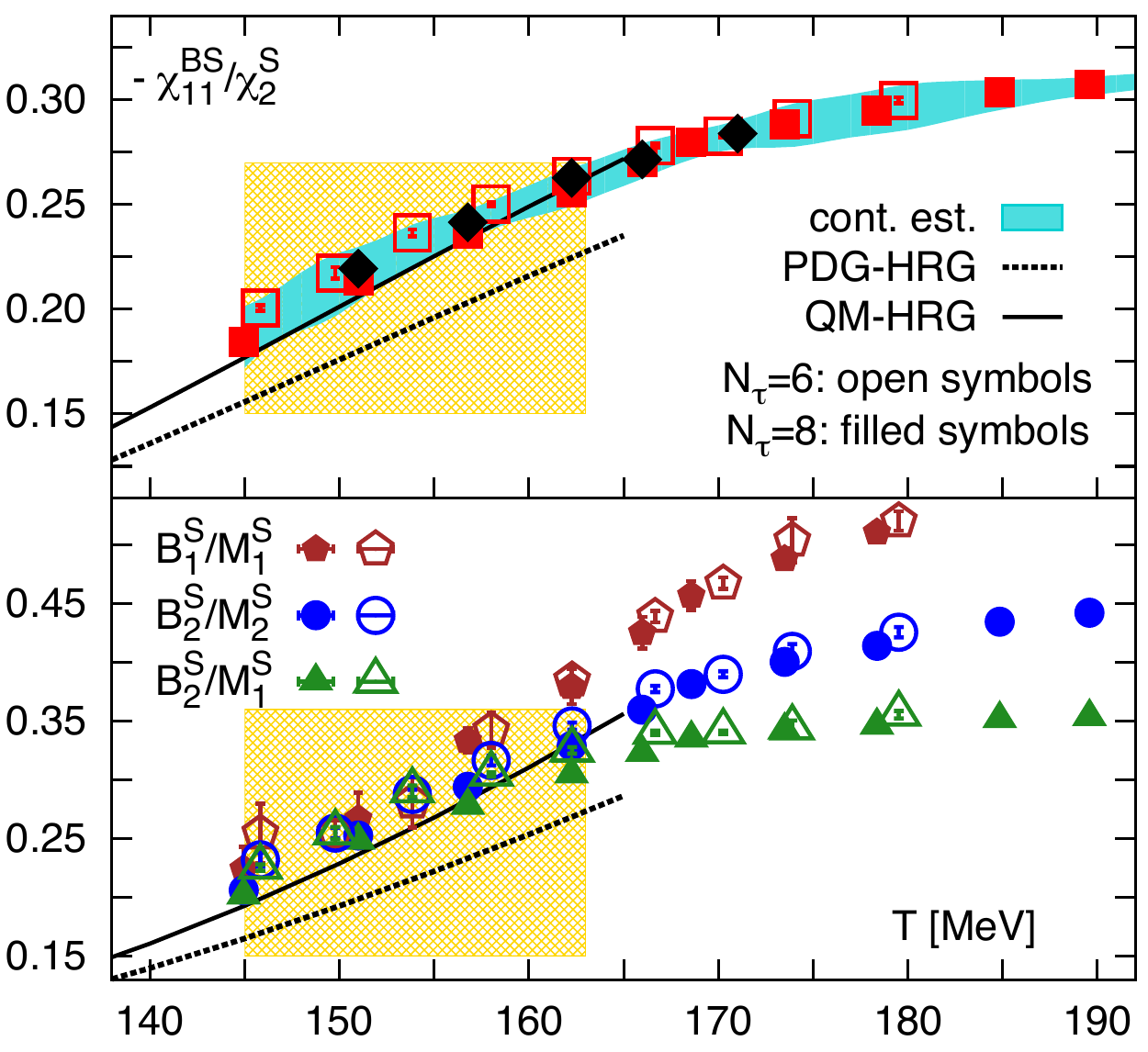}\includegraphics[width=0.3\textwidth,height=4.2cm]{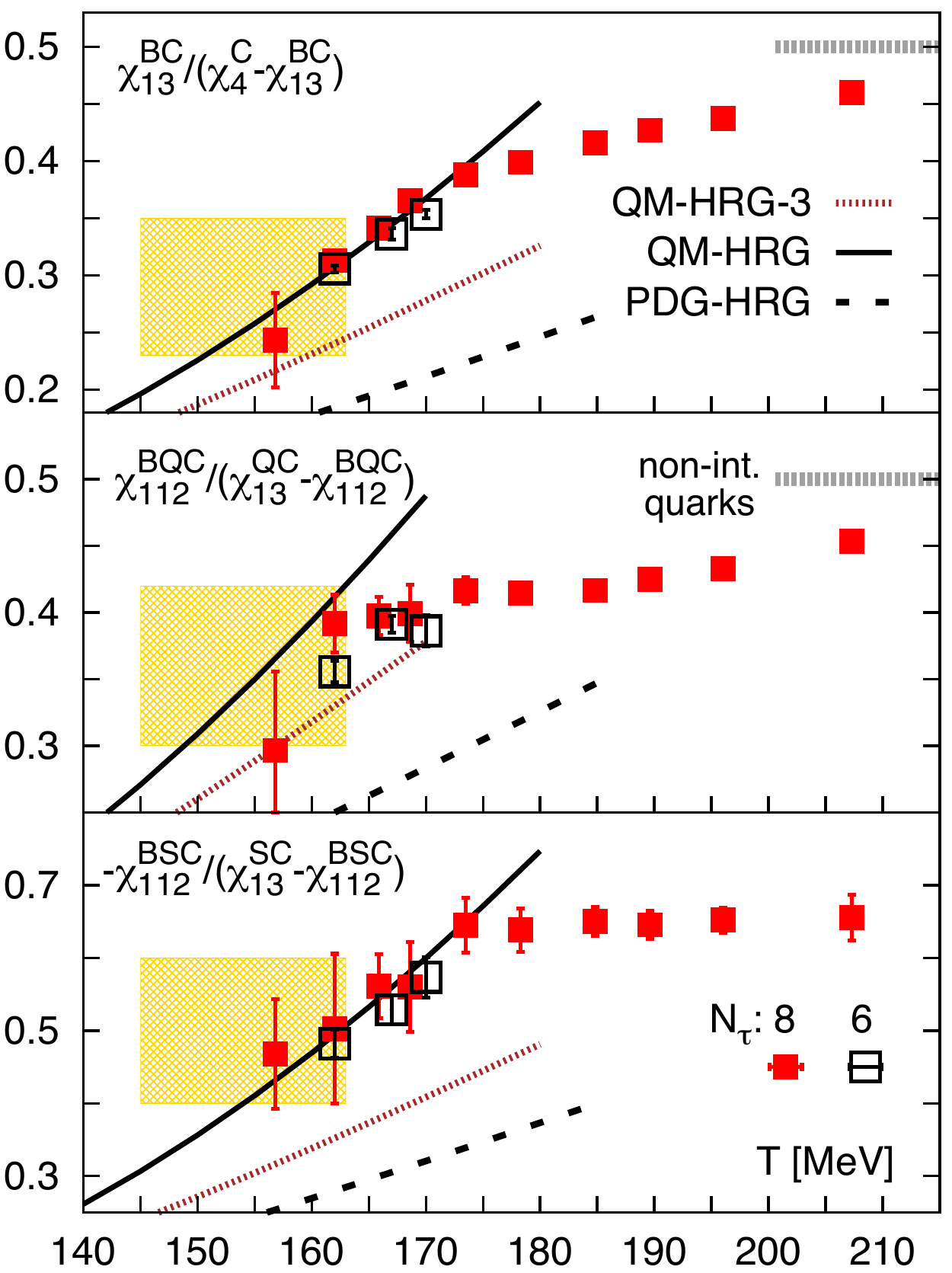}
\caption{The left plot shows the ratio of partial pressures of open strange hadrons $(P^{S,X}_{total})$, mesons $(P^{S,X}_M)$ and baryons $(P^{S,X}_B)$ calculated using Hadron Resonance Gas model with hadron spectrum 
from the particle data table (X=PDG) and from the relativistic Quark Model (X=QM). The plots in the middle and right panels show the
evidence of thermodynamic contributions from open strange and charm hadrons that are not listed in the PDG table but predicted in the relativistic Quark Model (QM), respectively.
Figures are taken from Refs.~\cite{MissingS,charmness}. Yellow bands shown in the middle and right plots label the chiral crossover temperature region.}
\label{fig:abundance}
\end{center}
\end{figure}

We start to look at the projected ratios of open strange hadrons in an uncorrelated hadron resonance gas with
the hadron spectrum from the particle data table and with additional hadron states predicted in
the relativistic Quark Model. As can be seen from the left plot in Fig.~\ref{fig:abundance} that additional
hadron states do contribute to the QCD thermodynamics~\footnote{The contents of hadron spectrum used in the HRG
does not come in the discussion in Section~\ref{sec:deconf} due to the fact that all the terms relevant for the details of
hadron spectrum canceled in the quantities discussed there.}. And it turns out that there are only small differences in partial meson pressures, 
however, larger differences in the baryon sector. This reflects the fact that more strange baryons are not listed in PDG compared to
open strange mesons. Based on this observation one can compute observables that reflect the ratio of partial pressures of 
open strange as well as charmed baryons to mesons on the lattice to search for the imprints of these additional hadrons in 
QCD thermodynamics.

Shown in the middle panel of Fig.~\ref{fig:abundance} are temperature dependences of 
ratios of baryon-strangeness correlation ($\chi_{11}^{BS}$) to quadratic strangeness fluctuations
$\chi_2^S$ and ratios of partial pressure of open strange mesons $M^S_i$ to strange baryons $B^S_i$.
Thus $-\chi_{11}^{BS}/\chi_2^S$ and $B_i^{S}/M_i^{S}$ reflect the relative contributions of strange baryons to 
open strange mesons in baryon-strangeness correlations and in partial pressures, respectively.
The temperature dependence of these quantities
in the chiral crossover region can be better described by the solid line 
(QM-HRG), i.e. results obtained from an HRG model using the hadron
spectrum predicted in the QM model. The commonly used HRG model based on the 
hadron spectrum listed in the PDG table, i.e. the dotted line (PDG-HRG) shown in the
plot, however, cannot describe the lattice data.

One can also investigate the charm sector in a similar way. Ratios of correlations and fluctuations, e.g.
$\chi_{13}^{BC}/(\chi_4^C-\chi_{13}^{BC})$, $\chi_{112}^{BQC}/(\chi_{13}^{QC}-\chi_{112}^{BQC})$ and $-\chi_{112}^{BSC}/(\chi_{13}^{SC}-\chi_{112}^{BSC})$
give the relative contributions of charmed baryons to open charm mesons, 
charged-charmed baryons to open charm charged mesons and 
strange-charmed baryons to 
strange-charmed mesons, respectively. Seen from the right panel of
Fig.~\ref{fig:abundance} same conclusion can be drawn as from the case in the strange sector,
i.e. the results obtained from the QM-HRG agree better with the lattice data than those from the PDG-HRG. 

The observations from Fig.~\ref{fig:abundance} serves as 
clear evidence for contributions from experimentally yet unobserved hadrons to 
the transition from hadronic matter to the Quark Gluon Plasma. 
The importance of these additional and non-PDG listed states 
has also been pointed out in Ref.~\cite{Majumder:2010ik}. 
In the next section we will discuss the influence of these additional states on 
the freeze-out conditions of strange hadrons in the heavy ion collisions.

\section{Freeze-out temperature of strange hadrons}
\begin{figure}[!th]
\begin{center} 
\includegraphics[width=0.39\textwidth]{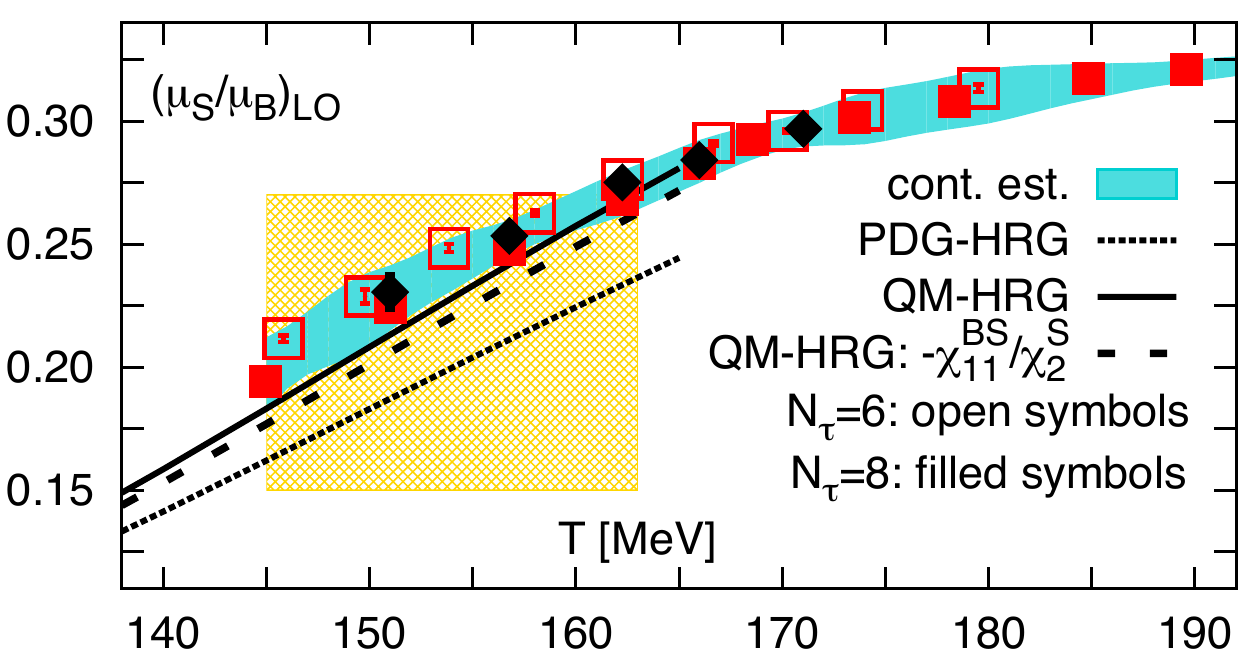}~\includegraphics[width=0.39\textwidth]{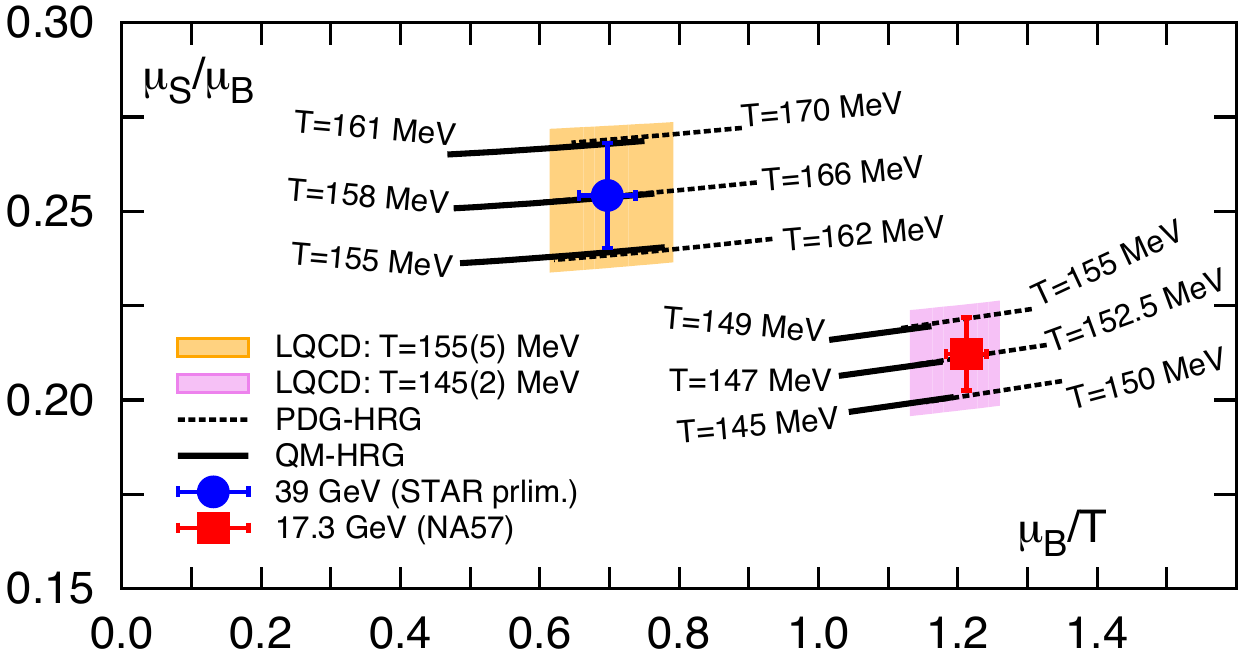}
\caption{Left: Temperature dependence of $\mu_S/\mu_B$ obtained from lattice QCD calculations and HRG model calculations using
hadron spectrum listed in the PDG table (PDG-HRG) and predicted from Quark Model (QM-HRG). Right: Comparison of freeze-out temperatures of strange hadrons
obtained by matching results obtained from lattice QCD (LQCD) and PDG-HRG as well as QM-HRG to the relation between $\mu_S/\mu_B$ and $\mu_B/T$ determined from STAR and NA57 experiments~\cite{HIC}. The temperatures obtained from QM-HRG are consistent with those from LQCD and are always lower than those from PDG-HRG by around 5$\sim$8 MeV. Figures are taken from Ref.~\cite{MissingS}.}
\label{fig:Tf}
\end{center}
\end{figure} 
The relation between $\mu_S/\mu_B$ and $\mu_B/T$ at the freeze-out can be extracted from heavy ion experiment data.
On the lattice we can calculate $\mu_S/\mu_B$ by using the Taylor expansion in $\mu_B/T$
\beq
\frac{\mu_S}{\mu_B}\simeq -\frac{\chi_{11}^{BS}}{\chi_2^S} - \frac{\chi_{11}^{QS}} {\chi_2^S} \frac{\mu_Q}{\mu_B} + \mathcal{O}(\mu_B^2).
\eeq
As shown in the left plot of Fig.~\ref{fig:Tf} ${\mu_S}/{\mu_B}$ is dominated by the term  $-\chi_{11}^{BS}/{\chi_2^S}$ which gets imprinted by
the additional strange hadron states. By varying the temperature ranges, one can match the relation between $\mu_S/\mu_B$ and $\mu_B/T$
obtained from lattice QCD (LQCD) and PDG-HRG as well as QM-HRG  to that between $\mu_S^f/\mu_B^f$ and $\mu_B^f/T^f$ extracted from the data measured in heavy ion experiments. The results are shown in
the right panel of Fig.~\ref{fig:Tf}. We found that the freeze-out temperature of strange hadrons lowers down by around 5-8 MeV when using lattice QCD
calculations compared to PDG-HRG and the results from QM-HRG agree with those from lattice QCD calculations. Thus the freeze-out temperature 
of strange hadrons becomes similar as the freeze-out temperature of light-quark hadrons.

\section{Summary}
By using the Highly Improved Staggered Quark (HISQ) action with a light to strange quark mass ratio of 1/20 and having charm 
quarks treated in the quenched approximation mainly on $N_\tau$=6 and 8 lattices
we have studied the thermodynamics of heavy-light hadron system by investigating on the 
ratios of cumulants of conserved net charge fluctuations. We found that the onset of dissociation of open strange 
and charmed hadron start in the chiral crossover temperature region.
We also found the indirect evidence of the existence of experimentally yet unobserved hadrons by comparing the lattice QCD
results with the predictions from QM-HRG and PDG-HRG on thermodynamic quantities. In the case of determining the freeze-out temperature
of strange hadrons one should take these experimentally yet unobserved hadron 
into account in the HRG models and consequently the freeze-out temperature of strange hadrons is brought down by around 5-8 MeV. 
This then leads to consistent freeze-out temperatures of strange hadrons with light-quark hadrons.

\end{document}